\documentclass[preprint,aps,showpacs,showkeys,12pt]{revtex4}
 \usepackage{graphicx}
\usepackage[pdftex]{color}
\usepackage{amsmath}
\usepackage{amssymb}
\usepackage{wasysym}

\newtheorem{theorem}{Theorem}[section]
\newtheorem{corollary}[theorem]{Corollary}
\newtheorem{proposition}[theorem]{Proposition}
\newtheorem{lemma}[theorem]{Lemma}
\newtheorem{definition}[theorem]{Definition}

\newtheorem{remark}[theorem]{Remark}

\def \beq{\begin{equation}}
\def \eeq{\end{equation}}
\def \beqa{\begin{eqnarray}}
\def \eeqa{\end{eqnarray}}
\def \beqan{\begin{eqnarray*}}
\def \eeqan{\end{eqnarray*}}
\def \bea{\begin{eqnarray}}
\def \eea{\end{eqnarray}}

\newcommand{\Hermitian}{{\mathfrak H}(n)}

\newcommand{\Density}{{\mathfrak D}(n)}
\newcommand{\support}{{\it supp}}
\newcommand{\supp}{{\it Supp}}
\def \um{\frac{1}{2}}

\newcommand{\tr}{\mathop{\rm tr}}

\newcommand{\Hi}{{\cal H}}
\newcommand{\R}{\mathbb{R}}

\newcommand{\N}{\mathbb{N}}

\newcommand{\Prob}{\mathbb{P}}
\newcommand{\D}{\mathbb{D}}
\newcommand{\E}{\mathbb{E}}
\newcommand{\Ec}{\mathcal{E}}
\newcommand{\Fc}{\mathcal{F}}
\newcommand{\Rc}{\mathcal{R}}

\newcommand{\qed}{\hfill $\Box$ \vskip 2ex}

\def \trace{\textrm{trace}}
\def \rank{\textrm{rank}}
\def \supp{\textrm{supp}}

\def \um{\frac{1}{2}}

\def \rank{\textrm{rank}}

\def \tr{\textrm{trace}}
\def \trace{\textrm{trace}}
\def \proof {\noindent {\em Proof. }}
\def \qed{\hfill $\Box$ \vskip 2ex}
\begin{document}
\title{Discrete-time classical and quantum  Markovian evolutions: Maximum entropy  problems on path space}\thanks{Work partially supported by the MIUR-PRIN Italian grant
``Identification and Control of Industrial Systems", by the GNAMPA-INDAM grant ``Teoria del Controllo per Sistemi Quantistici" and by the Department of Information Engineering research project ``QUINTET".}
\author{Michele Pavon}
 \homepage{http://www.math.unipd.it/~pavon}
\affiliation{Dipartimento di Matematica Pura ed Applicata, Universit\`a di Padova, via
Trieste 63, 35131 Padova, Italy}
\email{pavon@math.unipd.it}
\author{Francesco Ticozzi}
 \homepage{http://www.dei.unipd.it/~ticozzi}
\affiliation{Dipartimento di Ingegneria dell'Informazione,
Universit\`a di Padova, via Gradenigo 6/B, 35131 Padova, Italy}
\email{ticozzi@dei.unipd.it}

\date{\today}

\begin{abstract}
The theory of {\em Schr\"{o}dinger bridges} for diffusion processes is extended to classical and quantum discrete-time Markovian evolutions.  The solution of the path space maximum entropy problems is  obtained from  the {\em a priori} model in both cases via a suitable {\em multiplicative functional} transformation. In the quantum case, nonequilibrium {\em time reversal} of quantum channels is discussed and {\em space-time harmonic processes} are introduced. 
\end{abstract}

\pacs{05.30.-d,03.67.-a,03.67.Pp,02.50.-r} \keywords{Markov chain, quantum channel, maximum entropy problem,  Schr\"{o}dinger bridge, time reversal evolution, space-time harmonic function.} \maketitle

\section{Introduction}
In this paper, we study certain maximum entropy problems for discrete time and discrete state space Markov evolutions first considered by Erwin Schr\"{o}dinger in the early thirties for diffusion processes  \cite{S,S2}. In these problems, there is an {\em a priori} distribution on path space. Then new information becomes available in the form of the initial or terminal (or both) marginal distribution. One seeks a new path space distribution that has the correct marginal(s) and minimizes relative entropy from the prior distribution. Given the diffusion case results, their extension to Markov chains turns out to be rather straightforward. These results  serve us as a guideline and for comparison purposes in the more challenging quantum case.
 
To the best of our knowledge, the only previous discrete-time paper on this topic is \cite{B}, which deals  with the continuous state space case. The key results on Schr\"{o}dinger bridges, however, are there merely stated and the most delicate points in this extension, such as positivity of the space time harmonic function needed for the solution and existence and uniqueness of the latter, are altogether ignored. 
Here we show that the solution process can be obtained, in analogy to the diffusion case, via a suitable {\em multiplicative functional} transformation of the ``prior" Markov process, see Theorem \ref{maincl}. As in the diffusion case, an abstract result of Beurling and Jamison (Theorem \ref{BeJa}) can be used to prove existence and uniqueness of the solution of the Schr\"{o}dinger system for finite, irreducible and aperiodic Markov chains, see Corollary \ref{corexistence}. 

In order to derive corresponding results for quantum channels, we first need to develop various kinematical results. These concern extending the results on {\em time-reversal} of the channel by \cite{knill-nearoptimal,crooks}, and developing {\em space-time harmonic processes}. We also need to introduce a suitable concept of quantum ``trajectory": We consider a sequence of orthogonal projections selected from the spectral representations of a time-ordered sequence of observables. In spite of the obvious difficulties one can expect from the non commutative structure, we are actually able to solve two key maximum entropy problems on path space, cf. Theorems \ref{qtheo1}, \ref{qtheo2}. In the second case, the solution does not depend on the particular ``quantum path" chosen. Moreover, with the appropriate understanding of objects and properties, in both cases it bears a striking similarity to the classical case. 

The outline of the paper is as follows. In Section II, we outline the theory of  Schr\"{o}dinger bridges for diffusion processes. In Section III, as a warm up, we solve the simpler problems where only the initial or final marginal is prescribed. These are later needed for the purpose of comparison in the quantum case.  The  problem  where both marginals are fixed is solved in Section IV. Section V is devoted to introduce a class of Markovian processes from statistical quantum mechanics. In particular, their time reversal and space-time harmonic evolutions are there considered. In Section VI, after introducing a suitable concept of path, two maximum entropy problems are discussed and solved. The paper concludes with some final remarks  and an outlook on potential developments in Section VII.

\section{Classical Schr\"{o}dinger's bridges}
In 1931/32, Erwin Schr\"{o}dinger studied the following problem \cite{S,S2}. Consider the evolution of a cloud of $N$ independent Brownian particles. This cloud of particles has been observed having
at some initial time $t_0$ an empirical distribution equal to $\rho_0(x)dx$. At some later time $t_1$, an empirical distribution equal to  $\rho_1(x)dx$ is observed which considerably differs from what it should be according to the law of large numbers ($N$ is large, say of the order of Avogadro's number), namely
$$\left(\int_{t_0}^{t_1}p(t_0,y,t_1,x)\rho_0(y)dy\right)dx,
$$
where
$$p(s,y,t,x)=\left[2\pi(t-s)\right]
^{-\frac{n}{2}}\exp\left[-\frac{|x-y|^2} {2(t-s)}\right],\quad s<t
$$
is the transition density of the Wiener process. It is apparent that the particles have been transported in an unlikely way. But of the many unlikely ways in which this could have happened, which one is
the most likely? 

Schr\"{o}dinger considered first the simpler problem of the {\em Brownian bridge}, namely the case where the endpoint distributions are Dirac measures. He  introduced the three times (two-sided) transition density
$$q(s,x;t,y;u,z)=\frac{p(s,x;t,y)p(t,y;u,z)}{p(s,x;u,z)},\quad s<t<u,
$$
which the ``prior" and solution measures on path space must share. He then considered joint distributions of the initial and final time of the form
\begin{equation}\label{endpt} \mu(B)=\int_B p(t_0,x;t_1,y)d(\nu_0\times\nu_1)(x,y),
\end{equation}
where $B$ is a Borel set in $\R^n\times\R^n$, and $\nu_0$ and $\nu_1$ are $\sigma$-finite measures on $\R^n$. This form of the endpoint joint distribution, as proven by Jamison in \cite{Jam}, characterized derived processes that have the Markov property. General endpoint measures lead to {\em reciprocal processes} (one-dimensional Markov fields) introduced by Bernstein in \cite{BER}. 
Using then a  coarse graining approach on the initial and final state typical of Boltzmann and Lagrange multipliers, Schr\"{o}dinger went ahead to compute  the most likely endpoint distribution in the class (\ref{endpt}) under the ``prior" $p(s,x,t,y)$ and with the prescribed marginals. It turned out that the solution, namely the bridge from $\rho_0$ to $\rho_1$ over Brownian motion, has at each time a density $q$ that factors as $q(x,t)=\varphi(x,t)\hat{\varphi}(x,t)$, where $\varphi$ and $\hat{\varphi}$ are a $p$-harmonic and a $p$-coharmonic
functions, respectively. That is
\begin{eqnarray}\label{SY1}
&&\varphi(t,x)=\int
p(t,x,t_1,y)\varphi(t_1,y)dy,\\&&\hat{\varphi}(t,x)=\int
p(t_0,y,t,x)\hat{\varphi}(t_0,y)dy.\label{SY2}
\end{eqnarray}
The existence and uniqueness of a pair $(\varphi,\hat{\varphi})$ satisfying
(\ref{SY1})-(\ref{SY2}) and the boundary conditions
$\varphi(x,t_0)\hat{\varphi}(x,t_0)=\rho_0(x)$,
$\varphi(x,t_1)\hat{\varphi}(x,t_1)=\rho_1(x)$ was guessed by Schr\"{o}dinger
on the basis of his intuition. He was later shown to be quite right in
various degrees of generality by Fortet \cite{For}, Beurling \cite{Beu},
Jamison \cite{Jam}, F\"{o}llmer \cite{F2}. Jamison also showed \cite{Jam2} that the solution is a Markov diffusion process with drift given by $b(x,t)=\nabla\log\varphi$, i.e. its generator is ${\cal L}=\nabla\log\varphi\cdot\nabla+\frac{1}{2}\Delta$. It is namely an $h$-path process in the sense of Doob \cite{doob}.
Similar results may be obtained when the ``reference" process is any finite-energy Markovian diffusion possibly with creation and killing, see e.g. \cite{Wak} which is a survey paper with an extended bibliography. It took more than fifty years before F\"{o}llmer observed in \cite{F2} that all of this can be cast in the framework of a problem of {\em large deviations of the empirical distribution} on path space \cite{ellis,DS,DZ}, see also Appendix \ref{LD}. 




\section{Maximum entropy problems for Markov chains}

\subsection{Kinematics of Markov chains}\label{markovchains}
Consider a {\em Markov chain} $X=\{X(t); t=0,1,2,\ldots\}$ taking values in the finite or countably infinite set ${\cal X}$. Since ${\cal X}$ is countable, we can identify ${\cal X}$ with a subset of $\N$. Let us introduce the distribution of $X(t)$ given by $p_i(t)=\Prob(X(t)=i)$ and the {\em transition probabilities} $p_{ij}(t):=\Prob(X(t+1)=j|X(t)=i)$. They are connected through
\begin{equation}\label{FokkerP}p_j(t+1)=\sum_{i}p_{ij}(t)p_i(t).
\end{equation}
Let us agree that $\dag$ always indicates adjoint with respect to the natural inner product. Hence, in the case of matrices, it denotes transposition and, in the complex case below, transposition plus conjugation. We can then rewrite (\ref{FokkerP})  as $p(t+1)=P^\dag(t) p(t)$,
where $p(t)^\dag=(p_0(t),p_1(t),p_2(t),\ldots)$ and  $P(t)=(p_{ij}(t))$ is the transition matrix.
The latter is {\em stochastic}, i.e. all elements are nonnegative and  rows sum to one.
Let us introduce the {\em reverse-time transition probabilities}
\begin{equation}\label{revtimetransprob} q_{ji}(t,p(0)):=\Prob(X(t)=i|X(t+1)=j),
\end{equation}
where we have emphasized the dependence on the initial distribution $p(0)$. Also notice that the $q_{ji}$ depend on time even when the $p_{ij}$ don't. Let us find the relation between the $q_{ji}$ and the $p_{ij}$.
\begin{equation}\label{relpq}
p_i(t)p_{ij}(t)=\Prob(X(t)=i,X(t+1)=j)=p_j(t+1)q_{ji}(t,p(0)).
\end{equation}
Notice that for $p_j(t+1)=0$, $q_{ji}(t,p(0)), i\in\N$ may be defined arbitrarily any number between zero and one provided 
$$\sum_iq_{ji}(t,p(0))=1.
$$
Whenever $p_j(t+1)>0$, we get the relation
\begin{equation}\label{forbacktrans}
q_{ji}(t,p(0))=\frac{p_i(t)}{p_j(t+1)}p_{ij}(t).
\end{equation}
\begin{definition}A function $h:\N\times{\cal X}\rightarrow \R$ is called {\em space-time harmonic} for the transition mechanism $\{P(t);t=0,1,\ldots\}$ of a chain if, for every $t\ge 0$ and all $i,j\in{\cal X}$, it satisfies the {\em backward equation}
\begin{equation}h(t,i)=\sum_{j}p_{ij}(t)h(t+1,j).
\end{equation}
\end{definition}
Space-time harmonic functions, a terminology due to Doob and motivated by the case of diffusion processes, play a central role in constructing Schr\"{o}dinger bridges. They are closely related to a class of {\em martingales} that are  {\em instantaneous functions} of $X(t)$, see the Appendix \ref{dismart} for definition and properties.

\subsection{Simple path-space maximum entropy problems}
We consider first the simpler maximum entropy problems where only the initial or final marginals are prescribed. Although they turn out to be almost special cases of the initial-final marginal problem, they are important for statistical mechanics. 
\begin{definition}Let $p$ and $q$ be probability distributions on a finite or countably infinite set. We say that the {\em support} of $p$ is contained in the support of $q$ if $q_i=0\Rightarrow p_i=0$ and write $\support (p)\subseteq \support (q)$. The {\em Information Divergence} or {\em Relative Entropy}  or {\em Kullback-Leibler Index} of $q$ from $p$ is defined to be
\begin{equation}\label{KLdist}\D(p\|q)=\left\{\begin{array}{ll} \sum_ip(i)\log\frac{p(i)}{q(i)}, & \support (p)\subseteq \support (q),\\
+\infty , & \support (p)\not\subseteq \support (q).\end{array}\right.,
\end{equation} 
where, by definition, $0\cdot\log 0=0$.
\end{definition}
Let $X=\{X(0),X(1),\ldots\}$ be a Markov chain with state space ${\cal X}$, transition probabilities $(\pi_{ij}(t))$ and marginal probabilities $\Prob(X(t)=i)=\pi_i(t)$. Let ${\bf \Pi}$ denote the corresponding joint distribution of $\{X(0),X(1),\ldots, X(T)\}$ (distributions  on ${\cal X}^{T+1}$ are always denoted by capital, boldface letters). Let ${\cal D}(0,T;p^1)$ denote the family of Markovian distributions ${\bf P}$ on ${\cal X}^{T+1}$ that have marginal $p^1$  at time $T$, respectively, and have support contained in the support of ${\bf \Pi}$. Consider   the \\{\em Maximum Entropy Problem} (MEP1): 
\begin{equation}\label{MEs1}{\rm minimize}\quad \left\{\D({\bf P}\|{\bf \Pi}); {\bf P}\in{\cal D}(0,T;p^1)\right\}.
\end{equation}
Observe that the constraint on the final marginal in Problem (MEP1), may be expressed as a {\em linear} constraint on ${\bf P}$ as
$$\sum_{i_0}\sum_{i_1}\cdots\sum_{i_{T-1}}{\bf P}(i_0,i_1,\ldots,i_{T-1},i_T)=p^1(i_T).
$$
Thus Problem (MEP1) has the standard form of a maximum entropy problem \cite{cover}. Observe moreover that {\em existence} of a minimum point is always guaranteed since we are minimizing the lower semicontinuous function $\D(\cdot\|{\bf \Pi})$ with compact level sets \cite{ellis} over the closed set ${\cal D}(0,T;p^1)$. The same applies to the other maximum entropy problems considered below.

In order to facilitate the solution of Problem (MEP1) in (\ref{MEs1}), let us introduce the reverse time transition probabilities $q_{ji}(t)$ (\ref{revtimetransprob}) corresponding to ${\bf P}$ and $q^\pi_{ji}(t)$ corresponding to ${\bf \Pi}$. By the Markov property, we have
\begin{eqnarray}\label{§}{\bf P}(i_0,i_1,\ldots,i_{T-1},i_T)=q_{i_{1}i_0}(0)\cdot q_{i_{2}i_{1}}(1)\cdot q_{i_Ti_{T-1}}(T-1)\cdot p^1_{i_T},\\
{\bf \Pi}(i_0,i_1,\ldots,i_{T-1},i_T)=q^\pi_{i_{1}i_0}(0)\cdot q^\pi_{i_{2}i_{1}}(1)\cdot q^\pi_{i_Ti_{T-1}}(T-1)\cdot \pi_{i_T}(T).\label{backrec2}
\end{eqnarray}
We then get the following  representation for $\D({\bf P}\|{\bf \Pi})$.
\begin{lemma} \label{lemmarelentr} In the previous notation, assume that the support of ${\bf P}$ is contained in the support of ${\bf \Pi}$. Let $p(t)$ and $\pi(t)$ denote the marginals of ${\bf P}$ and ${\bf \Pi}$ at time $t$, respectively. Then
\begin{equation}\label{relentrdec}
\D({\bf P}\|{\bf \Pi})=\sum_{k=1}^T\sum_{i_k}\D\left(q_{i_ki_{k-1}}(k-1)\|q^\pi_{i_ki_{k-1}}(k-1)\right)p_{i_k}(k)+\D(p^1\|\pi(T)).
\end{equation}
\end{lemma}
The elementary proof is deferred to the Appendix \ref{prooflemma}.
\begin{theorem}\label{thmminentr} A solution to (MEP1) (\ref{MEs1}) is given by the distribution $\hat{{\bf P}}$ corresponding to the Markov chain with marginal distribution $p^1$ at time $T$ and reverse time transition mechanism equal to that of ${\bf \Pi}$, namely
\begin{equation}\label{qoptimcond} \hat{q}_{i_ki_{k-1}}(k-1)=q^\pi_{i_ki_{k-1}}(k-1), \quad k=1,2,\ldots, T.
\end{equation}
\end{theorem}
\noindent
\proof
Since both terms in (\ref{relentrdec}) are nonnegative, and $\D(p^1\|\pi(T))$ is invariant over ${\cal D}(p^1)$, the best we can hope for, when minimizing $\D({\bf P}\|{\bf \Pi})$, is to make the first term equal to zero. This is the case if (\ref{qoptimcond}) holds true.
\qed
\begin{remark}A moment thought reveals that the argument can be readily extended to show that $\hat{{\bf P}}$ is optimal in the larger class of all distributions on ${\cal X}^{T+1}$ that have marginal $p^1$  at time $T$. Indeed, it suffices in Lemma \ref{lemmarelentr}  to replace forward and reverse-time transition probabilities with conditional probabilities given the whole past or whole future trajectory, respectively. The Markovian nature of the solution $\hat{{\bf P}}$ therefore emerges as a by-product result of the variational problem. The same observation applies to the maximum entropy problems (MEP2) and (MEP3) below.
\end{remark}
Let us  compute the forward transition probabilities of $\hat{{\bf P}}$. Let $\hat{p}(t)$ and $\hat{p}_{ij}$ denote the marginal  at time $t$ and the forward transition probabilities of $\hat{{\bf P}}$, respectively. Let $\pi_{ij}$ be the forward transition probabilities of ${\bf \Pi}$. By (\ref{relpq}), we have
\begin{equation}\label{forwardbackward}\hat{p}_{i}(t)\hat{p}_{ij}(t)=\hat{p}_{j}(t+1)\hat{q}_{ji}(t),\quad \pi_i(t)\pi_{ij}=\pi_j(t+1)q^\pi_{ji}(t).
\end{equation}
Assume now that $\pi_i(t)>0, \hat{p}_i(t)>0, \forall i,  0\le t\le T$. Then (\ref{forwardbackward}) and (\ref{qoptimcond}) yield
\begin{equation}\label{multiplfunct}
\hat{p}_{ij}=\frac{\varphi(t+1,j)}{\varphi(t,i)}p_{ij}, \quad \varphi(t,i):=\frac{\hat{p}_i(t)}{\pi_i(t)}.
\end{equation}
Observe, moreover, that $\varphi$ is space-time harmonic with respect to the transition mechanism of ${\bf \Pi}$ since, by (\ref{multiplfunct}),
$$\sum_jp_{ij}\varphi(t+1,j)=\sum_j\hat{p}_{ij}\varphi(t,i)=\varphi(t,i).
$$
Hence, the solution is obtained from the reference probability ${\bf \Pi}$ through a ``multiplicative functional" transformation in analogy to the diffusion case \cite{Jam2}.
Consider now the case where the initial marginal is fixed. Let $p_{ij}(k)$ denote the transition probabilities of ${\bf P}$. Using 
\begin{equation}\label{multtransition}
\Prob(X(0)=i_0,X(1)=i_1,\ldots,X(n)=i_n)=p^0_{i_0}\cdot p_{i_oi_1}(0)\cdots p_{i_{n-2}i_{n-1}}(n-2)\cdot p_{i_{n-1}i_n}(n-1),
\end{equation}
and the corresponding one for ${\bf \Pi}$, we get along the lines of Lemma \ref{lemmarelentr} the representation of $\D({\bf P}\|{\bf \Pi})$:
\begin{equation}\label{relentrdec2}
\D({\bf P}\|{\bf \Pi})=\D(p^0\|\pi(0))+\sum_{k=0}^{T-1}\sum_{i_k}\D(p_{i_ki_{k+1}}(k)\|\pi_{i_ki_{k+1}}(k))p_{i_k}(k).
\end{equation}
We then get the following result.
\begin{theorem}\label{thmminentr2}A solution to the problem of maximizing $\left\{\D({\bf P}\|{\bf \Pi}); {\bf P}\in{\cal D}(0,T;p^0)\right\}$ (MEP2) is given by the distribution $\hat{{\bf P}}$ corresponding to the Markov chain with marginal distribution $p^0$ at time $0$ and forward transition mechanism equal to that of ${\bf \Pi}$, namely
\begin{equation}\label{qoptimcond2} \hat{p}_{ij}(t)=\pi_{ij}(t), \quad t=0,1,\ldots, T.
\end{equation}
\end{theorem}
Let us  compute the reverse-time transition probabilities of $\hat{{\bf P}}$. Let $\hat{p}(t)$ and $\hat{q}_{ji}(t)$ denote the marginal  at time $t$ and the reverse-time transition probabilities of $\hat{{\bf P}}$, respectively. Let $q^\pi_{ji}(t)$ be the reverse-time transition probabilities of ${\bf \Pi}$. By (\ref{relpq}), we have
\begin{equation}\label{forwardbackward2}\hat{p}_{j}(t+1)\hat{q}_{ji}(t)=\hat{p}_{i}(t)\pi_{ij}(t),\quad \pi_i(t)\pi_{ij}=\pi_j(t+1)q^\pi_{ji}(t).
\end{equation}
Assume now that $\pi_i(t)>0, \hat{p}_i(t)>0, \forall i,  0\le t\le T$. Then (\ref{forwardbackward2}) and (\ref{qoptimcond2}) yield
\begin{equation}\label{multiplfunct1}
\hat{q}_{ji}=\frac{\theta(t,i)}{\theta(t+1,j)}q^\pi_{ji}(t)., \quad \theta(t,i):=\frac{\hat{p}_i(t)}{\pi_i(t)}.
\end{equation}
Observe, moreover, that $\theta$ is space-time harmonic with respect to the reverse-time  transition mechanism of ${\bf \Pi}$ since, by (\ref{multiplfunct1}),
\begin{equation}\label{rtspacetime}\sum_iq^\pi_{ji}(t)\theta(t,i)=\sum_i\frac{\pi_i(t)}{\pi_j(t+1)}\pi_{ij}(t)\frac{\hat{p}_i(t)}{\pi_i(t)}=\theta(t+1,j).
\end{equation}
Again the solution is obtained from the reference probability ${\bf \Pi}$ through a ``multiplicative functional" transformation. The special case  of property (\ref{rtspacetime}) where $\pi_{ij}$ do not depend on time and $\hat{p}(t)\equiv\bar{\pi}$ a stationary distribution is strictly connected to a strong form of the second law and a principle of minimum dissipation \cite{PT}. We finally mention that Problems (MEP1)-(MEP2) are connected to {\em optimal stochastic control} problems \cite{pavwak} (corresponding results for continuous time Markov chains are contained in an unpublished 1988 Master Thesis by C. M. Pelaggi \cite{pelaggi}).

\section{Schr\"{o}dinger bridges for Markov chains}
\subsection{The initial-final marginal problem}
In the same notation as in the previous section, let ${\cal D}(0,T;p^0,p^1)$ denote the family of Markovian distributions ${\bf P}$ on ${\cal X}^{T+1}$ that have marginals $p^0$ at time $0$ and $p^1$  at time $T$, respectively, and have support contained in the support of ${\bf \Pi}$. We consider   the following \\ {\em Maximum Entropy Problem} (MEP3): 
\begin{equation}\label{ME3}{\rm minimize}\quad \left\{\D({\bf P}\|{\bf \Pi}); {\bf P}\in{\cal D}(0,T;p^0,p^1)\right\}.
\end{equation}
Consider the representation (\ref{relentrdec2}). Since $\D(p^0\|\pi(0))$ is invariant over ${\cal D}(0,T;p^0,p^1)$, Problem (MEP3) in (\ref{ME3}) is equivalent to
\begin{equation}\label{ME4}{\rm minimize}\quad \left\{\sum_{k=0}^{T-1}\sum_{i_k}\D(p_{i_ki_{k+1}}(k)\|\pi_{i_ki_{k+1}}(k))p_{i_k}(k); {\bf P}\in{\cal D}(0,T;p^0,p^1)\right\}.
\end{equation}
Rather than following the standard Schr\"{o}dinger-F\"{o}llmer approach as in \cite{B}, we develop a discrete  counterpart of the approach of Nagasawa for diffusion processes \cite{nag} which relies heavily on Girsanov's theory \cite{KS}. Consider  a {\em space-time harmonic function} $\varphi$ for the {\em reference} stochastic evolution ${\bf \Pi}$, namely
\begin{equation}\label{spacetimeharmonic}\varphi(t,i)=\sum_{j}\pi_{ij}(t)\varphi(t+1,j), \quad 0\le t \le(T-1).
\end{equation}
Assume that $\varphi(t,i)>0, \;\forall 0\le t\le T, \forall i\in{\cal X}$. Then Problem (\ref{ME4}) is equivalent to
\begin{equation}\label{ME5}{\rm minimize}\left\{J({\bf P}); {\bf P}\in{\cal D}(0,T;p^0,p^1)\right\},
\end{equation}
where
$$J({\bf P})=\sum_{k=0}^{T-1}\sum_{i_k}\D(p_{i_ki_{k+1}}(k)\|\pi_{i_ki_{k+1}}(k))p_{i_k}(k)+\sum_{i_0}\log\varphi (0,i_0)p^0_{i_0}-\sum_{i_T}\log\varphi(T,i_T)p^1_{i_T},
$$
since the last two terms are invariant over ${\cal D}(0,T;p^0,p^1)$.
Next, observe that
$$\sum_{i_0}\log\varphi (0,i_0)p^0_{i_0}-\sum_{i_T}\log\varphi(T,i_T)p^1_{i_T}=\sum_{k=0}^{T-1}\left[\sum_{i_k}\log \varphi(k,i_k)p_{i_k}(k)-\sum_{i_{k+1}}\log \varphi(k+1,i_{k+1})p_{i_{k+1}}(k+1)\right].
$$
Moreover,
\begin{eqnarray}\nonumber\sum_{i_k}\log \varphi(k,i_k)p_{i_k}(k)-\sum_{i_{k+1}}\log \varphi(k+1,i_{k+1})p_{i_{k+1}}(k+1)\\=\sum_{i_k}\log \varphi(k,i_k)p_{i_k}(k)-\sum_{i_{k+1}}\log \varphi(k+1,i_{k+1})\sum_{i_k}p_{i_ki_{k+1}}(k)p_{i_{k}}(k)\nonumber\\\nonumber=\sum_{i_k}\sum_{i_{k+1}}p_{i_ki_{k+1}}(k)\log \varphi(k,i_k)p_{i_k}(k)-\sum_{i_{k+1}}\log \varphi(k+1,i_{k+1})\sum_{i_k}p_{i_ki_{k+1}}(k)p_{i_{k}}(k)\\\nonumber=\sum_{i_k}\sum_{i_{k+1}}p_{i_ki_{k+1}}(k)\left[\log \varphi(k,i_k)-\log \varphi(k+1,i_{k+1})\right]p_{i_k}(k).
\end{eqnarray}
Hence, $J({\bf P})$ may be rewritten as
$$J({\bf P})=\sum_{k=0}^{T-1}\sum_{i_k}\D\left(p_{i_{k}i_{k+1}}(k)\|\pi_{i_{k}i_{k+1}}(k)\frac{\varphi(k+1,i_{k+1})}{\varphi(k,i_k)}\right)p_{i_{k}}(k).
$$
Define
\begin{equation}\label{optimaltransprob}\hat{p}_{ij}(k)=\pi_{ij}(k)\frac{\varphi(k+1,j)}{\varphi(k,i)}.
\end{equation}
Notice that $\hat{p}_{ij}\ge 0$ and, by (\ref{spacetimeharmonic}),
$$\sum_j \hat{p}_{ij}(k)=\sum_j\pi_{ij}(k)\frac{\varphi(k+1,j)}{\varphi(k,i)}=\frac{\varphi(k,i)}{\varphi(k,i)}=1.
$$
Consider the probabilities $\hat{p}(t)$ defined by the recursion
$$\hat{p}_j(t+1)=\sum_i\hat{p}_{ij}(t)\hat{p}_i(t),\quad \hat{p}_i(0)=p^0_i.
$$
Define
$$\hat{\varphi}(t,i):=\frac{\hat{p}_i(t)}{\varphi(t,i)}.
$$
We get 
\begin{equation}
\hat{\varphi}(t+1,j)=\frac{p_j(t+1)}{\varphi(t+1,j)}=\frac{\sum_i\pi_{ij}(t)\frac{\varphi(t+1,j)}{\varphi(t,i)}(t)p_i(t)}{\varphi(t+1,j)}=\sum_i\pi_{ij}(t)\frac{p_i(t)}{\varphi(t,i)}=\sum_i\pi_{ij}(t)\hat{\varphi}(t,i),
\end{equation}
namely $\hat{\varphi}(t,i)$ is {\em space-time co-harmonic}. We have therefore established the following theorem:
\begin{theorem} \label{maincl}Suppose there exists a pair of nonnegative functions $(\varphi,\hat{\varphi})$ defined  on $[0,T]\times{\cal X}$ and satisfying the system
\begin{eqnarray}\label{scsi1}\varphi(t,i)=\sum_{j}\pi_{ij}(t)\varphi(t+1,j), \quad 0\le t \le(T-1)\\
\hat{\varphi}(t+1,j)=\sum_i\pi_{ij}(t)\hat{\varphi}(t,i),\quad 0\le t \le(T-1),\label{scsi2}
\end{eqnarray}
as well as the boundary conditions
\begin{equation}\label{scsi3}
\varphi(0,i)\cdot\hat{\varphi}(0,i):=p^0_i,\quad \varphi(T,i)\cdot\hat{\varphi}(T,i):=p^1_i,\quad \forall i\in{\cal X}.
\end{equation}
Suppose moreover that $\varphi(t,i)>0,\; \forall 0\le t\le T, \forall i\in{\cal X}$. Then, the Markov distribution $\hat{{\bf P}}$ in ${\cal D}(0,T;p^0,p^1)$ having transition probabilities
\begin{equation}\hat{p}_{ij}(t)=\pi_{ij}(t)\frac{\varphi(t+1,j)}{\varphi(t,i)}
\end{equation}
solves problem (MEP3) \eqref{ME3}.
\end{theorem}
Notice that if $(\varphi,\hat{\varphi})$ satisfy (\ref{scsi1})-(\ref{scsi2})-(\ref{scsi3}), so does the pair $(c\varphi,\frac{1}{c}\hat{\varphi})$ for all $c>0$. Hence, uniqueness for the Schr\"{o}dinger system is always intended up to such multiplications.
As in the diffusion case, the problem is now reduced to establish, under suitable assumptions, existence and uniqueness for the Schr\"{o}dinger system (\ref{scsi1})-(\ref{scsi2})-(\ref{scsi3}) (notice that this issue is not even mentioned in \cite{B}). Also observe that the solution of (MEP1) and (MEP2) may be viewed as a special cases of the solution of (MEP3) where the role of the space-time co-harmonic function $\hat{\varphi}(t,i)$ is played by the prior probabilities $\pi_i(t)$. Indeed, let us  compute the reverse-time transition probabilities $\hat{q}_{ji}(t)$ of $\hat{{\bf P}}$. By (\ref{relpq}) and using $\hat{p}_i(t)=\varphi(t,i)\cdot\hat{\varphi}(t,i)$, we get
\begin{eqnarray}\nonumber\hat{q}_{ji}(t)=\frac{\hat{p}_{i}(t)}{\hat{p}_j(t+1)}\hat{p}_{ij}(t)=\frac{\hat{p}_{i}(t)}{\hat{p}_j(t+1)}\pi_{ij}(t)\frac{\varphi(t+1,j)}{\varphi(t,i)}\\=\pi_{ij}(t)\frac{\hat{\varphi}(t,i)}{\hat{\varphi}(t+1,j)}=q^\pi_{ji}(t)\frac{\xi(t,i)}{\xi(t+1,j)},\nonumber
\end{eqnarray}
where
$$\xi(t,i):=\frac{\hat{\varphi}(t,i)}{\pi_i(t)}.
$$
Moreover, 
$$\sum_iq^\pi_{ji}(t)\xi(t,i)=\sum_i\frac{\pi_i(t)}{\pi_j(t+1)}\pi_{ij}(t)\frac{\hat{\varphi}(t,i)}{\pi_i(t)}=\xi(t+1,j).
$$
namely, $\xi$ is space-time harmonic with respect to the reverse-time  transition mechanism of ${\bf \Pi}$. We then get the suggestive formula
$$\hat{p}_i(t)=\varphi(t,i)\cdot\xi(t,i)\cdot\pi_i(t),
$$
showing that the ``a posterior" marginals are obtained from the ``a priori" marginals through multiplication by a space-time harmonic function for the forward transition mechanism and by a space-time harmonic function for the reverse time transition mechanism. In problems (MEP1) and (MEP2), we have $\xi(t,i)\equiv 1$ and $\varphi(t,i)\equiv 1$, respectively.  Namely, the solutions to (MEP1) and (MEP2) appear as particular cases of the solution to (MEP3), where only the forward or only the backward transition mechanism undergoes a multiplicative functional transformation induced by a space-time harmonic function, whereas in (MEP3) {\em both} transition mechanisms are subject to such a transformation. Finally, notice that if we exchange $p^0$ and $p^1$, the solution will simply be the time reversal of $\hat{{\bf P}}$: This intrinsic time reversibility of the bridge attracted the interest of Schr\"{o}dinger and Kolmogorov, see the titles of \cite{S,kolmogorov}.

\subsection{Existence and uniqueness for the Schr\"{o}dinger system}
Existence and uniqueness of the solution to the Schr\"{o}dinger system (\ref{scsi1})-(\ref{scsi2})-(\ref{scsi3}) follows from a very deep result of Beurling \cite{Beu}, suitably extended by Jamison \cite[Theorem 3.2]{Jam}.
\begin{theorem}\label{BeJa} Let S be a $\sigma$-compact metric space. Let $\mu_1$ and $\mu_2$ be probability measures on its $\sigma$-field $\Sigma$ of Borel sets. Suppose $p$ is a continuous, strictly positive function on $S\times S$. Then there exists a unique pair $\mu$, $\pi$ of measures on $S\times S$ such that
\begin{enumerate}
\item $\mu$ is a probability measure and $\pi$ is a $\sigma$-finite measure;
\item $\mu(E\times S)=\mu_1(E),\quad \mu(S\times E)=\mu_2(E),\quad E\in S$;
\item $\frac{d\mu}{d\pi}=p$.
\end{enumerate}
\end{theorem}
It is argued \cite[pp. 76-77]{Jam} that the existence and uniqueness of this theorem are equivalent to existence and uniqueness for an abstract Schr\"{o}dinger system where $\mu_1$ and $\mu_2$ play the role of initial and final marginals, respectively, and $p$ the role of the Markov transition density of the prior probability on path space. The fact that \cite{Jam} deals with the continuous time case is immaterial: Only the initial and final times are involved in this  result.

In order to apply the Beurling-Jamison theorem to our setting, we first need a few observations. In our setting, $S={\cal X}$ is finite or countably infinite. We endow ${\cal X}$ with the {\em discrete metric}, namely $d(x,y)=1$ for all pairs $(x,y)\in {\cal X}\times {\cal X}, x\neq y$, and $d(x,x)=0,\forall x\in{\cal X}$. This makes ${\cal X}$ a $\sigma$-compact metric space where the transition probability $p(s,x,t,y):=\Prob(X(t)=y|X(s)=x)$ is continuous on ${\cal X}\times {\cal X}$ for any pair of times $s<t$. The Borel $\sigma$-field of ${\cal X}$, denoted ${\cal B}({\cal X})$ is then simply the family of all subsets of ${\cal X}$. The probability density with respect to the counting measure $\lambda_X$ is simply the probability of that element in ${\cal X}$.
\begin{theorem}\label{existenceun}Let $X=\{X(0),X(1),\ldots\}$ be a Markov chain with state space ${\cal X}$ and transition probabilities $\pi_{ij}(t)$. Assume
\begin{enumerate}
\item $p^1$ is a distribution on ${\cal X}$ with $p^1_x>0, \forall x\in{\cal X}$;
\item $p(0,x,T,y)>0, \forall x,y\in{\cal X}$.
\end{enumerate}
Then the Schr\"{o}dinger system (\ref{scsi1})-(\ref{scsi2})-(\ref{scsi3}) has a unique solution with $\varphi(t,x)>0,\; \forall 0\le t\le T, \forall x\in{\cal X}$. 
\end{theorem}
\proof Existence and uniqueness for the Schr\"{o}dinger system follows from Theorem \ref{BeJa}. From (\ref{scsi3}), the nonnegativity of $\varphi$ and the assumption on $p^1$, it follows that $\varphi(T,x)>0, \forall x\in{\cal X}$. Recalling that
$$\sum_j\pi_{ij}(t)=1,\quad \forall t,
$$
we get from (\ref{scsi1}) that $\varphi(t,x)>0,\; \forall 0\le t\le T, \forall x\in{\cal X}$. 
\qed
In many important applications, the prior transition probabilities do not depend on time. We get the following result for finite, irreducible and aperiodic Markov chains.
\begin{corollary}\label{corexistence}Let $\{X(0),X(1),\ldots\}$ be a Markov chain with finite state space ${\cal X}$ and transition matrix $\Pi=(\pi_{ij})$. Assume
\begin{enumerate}
\item $p^1$ is a distribution on ${\cal X}$ with $p^1_x>0, \forall x\in{\cal X}$;
\item the matrix $P^T$ has all positive elements.
\end{enumerate}
Then the Schr\"{o}dinger system (\ref{scsi1})-(\ref{scsi2})-(\ref{scsi3}) has a unique solution with $\varphi(t,x)>0,\; \forall 0\le t\le T, \forall x\in{\cal X}$. 

\end{corollary}
The implications of these results for {\em large deviations} is outlined in the Appendix \ref{LD}.

\section{Quantum Markov channels}\label{quantum}

In the following sections, we show how the framework developed for classical Markov chains can be extended to the class of discrete-time, Markovian quantum processes generated by families of {\em quantum operations}. We begin by establishing some notation and recalling some basic facts from quantum statistical theory. Some classical references are \cite{vonneumann,sakurai,holevo}. A basic and clear exposition from the quantum information viewpoint can be found in \cite{nielsen-chuang}, while an excellent introduction to  quantum probability concepts and formalism is \cite{maassen-qp}. 

\subsection{Quantum probabilities, entropy and quantum operations}

Consider a finite-dimensional quantum system ${\cal Q},$ with associated Hilbert space $\Hi_{\cal Q}$ isomorphic to ${\mathbb C}^n.$ In the quantum probability formalism, random variables or {\em observables} for the system are represented by Hermitian matrices $X\in\Hermitian.$ They admit a spectral representation $X=\sum_j x_j \Pi_j,$ where each real eigenvalue $x_j$ represents the random outcome associated to the quantum {\em event} corresponding to the orthogonal projection $\Pi_j$.  The role of the probability distributions is played here by positive-definite, unit-trace matrices $\rho\geq 0, \trace(\rho)=1,$ called {\em density matrices}. The set $\Density$ of density matrices is convex and has the rank-one orthogonal projections as extreme points. 
Assume that the density matrix associated to the state of the system is $\rho$. The probability of measuring $x_j,$ or in general the probability associated to the quantum event $\Pi_j,$ is given by:
$$\Prob_\rho(\Pi_j)=\trace(\Pi_j\rho\Pi_j).$$
If the outcome corresponding to an event $\Pi_j$ has been measured, the density matrix conditioned on the measurement record is:
$$\rho _{|{\Pi_j}}=\frac{1}{\trace\Pi_j\rho\Pi_j}\Pi_j\rho\Pi_j.$$
Hence, the joint probability of obtaining $\Pi_k$  after $\Pi_j$ in subsequent measurements can be computed by:
\beq\label{joinp}\Prob_\rho(\Pi_j,\Pi_k)=\trace(\Pi_k\Pi_j\rho\Pi_j\Pi_k),\eeq
where the order of events is relevant. Similarly one obtains nested expressions for joint probabilities of arbitrary event sequences. Expectations are then computed by the trace functional: 
$$\E_\rho(X)=\sum_jx_j\trace(\Pi_j\rho\Pi_j)=\trace(\rho X).$$ Notice that this implies that if the measurement has occurred, but the outcome has not been recorded, the correct conditional density matrix is:
$$\rho _{|{X}}=\sum_j\frac{1}{\trace\Pi_j\rho\Pi_j}\Pi_j\rho\Pi_j\cdot\Prob_\rho(\Pi_j)=\sum_j\Pi_j\rho\Pi_j,$$
which is in general different from the pre-measurement $\rho,$ in contrast with the classical case. We refer to these ``blind'' measurement processes as {\em non-selective} measurements.

For any matrix $M$, the {\em support} of $M$, denoted $\support(M)$, is the orthogonal complement of $\ker(M)$. Given two density matrices $\rho,\sigma,$ the {\em quantum relative entropy} is defined by:
\begin{equation}\label{umegakirelentr}\D(\rho\|\sigma)=\left\{\begin{array}{ll} \tr(\rho(\log \rho -\log \sigma)), & \support (\rho)\subseteq \support (\sigma),\\
+\infty , & \support (\rho)\cap \support (\sigma)^\perp\neq 0\end{array}\right.,
\end{equation} 
As in the classical case, quantum relative entropy has the
property of a pseudo-distance (see e.g. \cite{nielsen-chuang, Ve}):
The Klein's Inequality $\D(\rho||\sigma)\geq 0$ holds, equality occurring if and only if
$\rho=\sigma$.
Moreover, quantum relative entropy is continuous where it is not
infinite and it is jointly convex, but not symmetric, in its
arguments.

A wide class of physically relevant, Markovian transition mechanisms are represented by linear, Trace Preserving and Completely Positive (TPCP) maps from density matrices to density matrices. A TPCP map $\Ec^\dag$, in turn, can be represented by a Kraus operator-sum \cite{kraus}, i.e.:  
$$\rho_{t+1}=\Ec^\dag(\rho_t)=\sum_j M_j\rho_t M^\dag_j,$$
where the $n\times n$ matrices $M_j$ must satisfy $\sum_j M_j^\dag M_j = I$ in order for $\Ec^\dag $ to be trace preserving. \footnote{ We employ the adjoint for maps acting on states to be consistent with the classical notation, where the transition matrix $P^\dag$ acts on  probability distributions while $P$ acts on functions, see  \cite{nelson-adjoint},\cite[Chapter 13]{N1} and \cite{PT} for a discussion on the role of duality relations for Markov evolutions.} Following common conventions in quantum information \cite{nielsen-chuang}, we refer to general Kraus operator-sums as Kraus maps, and to trace-non-increasing ones as {\em quantum operations}. 
As an example that will be used in the following Sections, notice that the non-selective measurement process $\Ec^\dag(\rho)=\sum_j\Pi_j\rho\Pi_j$ is a valid TPCP map, while the un-normalized conditioning $\Ec^\dag(\rho)=\Pi_j\rho\Pi_j$ is a quantum operation, since it is trace non-increasing. 

The action of the dynamics on observables can be derived by duality with respect to the Hilbert-Schmidt inner product:
$$\trace(X\Ec^\dag(\rho_t))=\trace(\Ec(X)\rho_t),$$
where $\Ec(X)=\sum_j M_j^\dag X M_j.$ It follows that if $\Ec^\dag(\cdot)$ is trace-preserving, then $\Ec(\cdot)$ is identity preserving and vice-versa.

We are now in a position to introduce a quantum analogue of {\em space-time harmonic functions}. Consider a quantum Markov process, generated by $\rho_0$ and a sequence of TPCP maps $\{\Ec^\dag_t\}_{t\in[0, T-1]}.$
\begin{definition} [Quantum space-time harmonic property] \label{qharmonic} A sequence of observables $\{Y_t\}_{t\in[0, T-1]}$ is said to be space-time harmonic with respect to the family  $\{\Ec_t\}_{t\in[0, T-1]}$ if:
\beq \label{qsth} Y_{t}=\Ec_t(Y_{t+1}). \eeq
\end{definition}
As in the classical case, space-time harmonic processes will be shown to play a central role in the solution of maximum entropy problems on path spaces.

\subsection{Time-reversal of quantum operations}\label{timerev}

Another key ingredient in the study of maximum entropy problems on path space, is, very much like for classical Markov chains,  the reverse-time transition mechanism. 
We give here an intuitive derivation of the time-reversal of a quantum operation through an analogy with the classical case, referring to \cite{PT} for a thorough discussion in the framework of adjoint Markov semigroups \cite{nelson-adjoint}, its connections to the equilibrium case \cite{crooks} and previous results from the {\em quantum error correction} literature \cite{knill-qec,knill-nearoptimal}. 

If we write $\rho_t=\sum_ip_i\Pi_{t,i},$ $\rho_{t+1}=\sum_jq_j\Pi_{t+1,j},$ we can think of the density matrices as classical probability densities over the abelian algebras of quantum events generated by their spectral families. The probability of measuring $\Pi_{j,t+1}$, after $\Pi_{i,t}$ has been measured at time $t$ and $\Ec^\dag$ has acted on the system, is given by: 
\beqa \Prob_{\Ec^\dag,\rho_t}(\Pi_{i,t},\Pi_{j,t+1})&=&\trace\left(\Pi_{j,t+1}\sum_kM_k \Pi_{i,t}\rho_t\Pi_{i,t} M_k^\dag \Pi_{j,t+1}\right),\nonumber\\
&=& \label{forwardp} \trace\left(\Pi_{j,t+1}\sum_kM_k \Pi_{i,t} M_k^\dag \Pi_{j,t+1}\right)p_i,
\eeqa
obtaining an analogous of the transition probabilities in the classical case. Following this analogy, the reverse-time $\Rc^\dag_{\Ec,\rho_t}$ is then required to satisfy:
\beq\Prob_{\Ec^\dag,\rho_t}(\Pi_{i,t},\Pi_{j,t+1})=\trace\left(\Pi_{j,t}\sum_kR_k(\Ec,\rho_t) \Pi_{i,t+1} R_k^\dag(\Ec,\rho_t) \Pi_{j,t}\right)q_j=\Prob_{\Rc^\dag_{\Ec,\rho_t},\rho_{t+1}}(\Pi_{j,t+1},\Pi_{i,t}).\label{backwardp}\eeq
Notice that \eqref{forwardp}-\eqref{backwardp} provide us with the equivalent of \eqref{relpq} for classical chains. Assume for now that $\rho_{t+1}$ is full-rank. From \eqref{forwardp}, using the cyclic property of the trace and the fact that $\rho_{t+1}$ and $\Pi_{j,t+1}$ commute for all $j$, we obtain:
\beqa \Prob_{\Ec^\dag,\rho_t}(\Pi_{i,t},\Pi_{j,t+1})&=&\sum_k\trace\left(\Pi_{j,t+1}M_k \Pi_{i,t}\rho_t\Pi_{i,t} M_k^\dag \Pi_{j,t+1}\right),\nonumber\\
&=&\sum_k  \trace\left(\Pi_{i,t}\rho_t^\um M^\dag_k\rho_{t+1}^{-\um} \Pi_{j,t+1} \rho_{t+1}\Pi_{j,t+1}\rho_{t+1}^{-\um}M_k \rho_t^\um\Pi_{i,t}\right),\nonumber\\
&=& \trace\left(\Pi_{i,t}\sum_k \left(\rho_t^\um M^\dag_k\rho_{t+1}^{-\um}\right) \Pi_{j,t+1} \left(\rho_{t+1}^{-\um}M_k \rho_t^\um\Pi_{i,t}\right)\right)q_j.\nonumber
\eeqa
Hence, we define 
\begin{equation}R_j(\Ec,\rho_t)=\rho^{-\um}_{t+1}M_j\rho_t^\um,
\end{equation} 
and a Kraus map: 
\beq\label{rev}\Rc^\dag_{\Ec,\rho_t}(\cdot)=\sum_jR_j(\Ec,\rho_t)(\cdot)R^\dag_j(\Ec,\rho_t).\eeq 

In \cite{PT}, we show that this map is in fact a quantum operation, that it can be augmented to a trace-preserving quantum operation, and that it is the correct time-reversal for $\Ec$ with respect to the initial density $\rho_t.$ This is established also in the case $\rank (\rho_{t+1})<n$, thereby extending the  results in \cite{knill-nearoptimal} (in this case, the inverse has to be replaced by the Moore-Penrose pseudoinverse, cf. \cite{horn-johnson}). 

For any $\rho$ and $\Ec^\dag$ with Kraus operators $\{M_k\},$ define the map ${\cal T}_{\rho}$ from {\em quantum operations to quantum operations} 
\beq {\cal T}_{\rho}:\Ec^\dag\mapsto {\cal T}_{\rho}(\Ec^\dag),\eeq
where ${\cal T}_{\rho}(\Ec^\dag)$ has Kraus operators $\{\rho^\um M^\dag_k (\Ec(\rho))^{-\um} \}.$ The results of  \cite{knill-nearoptimal} show that the action of ${\cal T}_{\rho}$ is independent of the particular Kraus representation of $\Ec^\dag$. With this definition, we have that $${\cal T}_{\rho_t}(\Ec^\dag)=\Rc^\dag_{\Ec,\rho_t}.$$

\begin{theorem} [Quantum Operation Time Reversal, \cite{PT}]\label{qtimerev} Let  $\Ec^\dag$ be a TPCP map. If $\rho_{t+1}=\Ec^\dag(\rho_t),$ then for any $\rho_t\in\mathfrak{D}(n),$ $\Rc^\dag_{\Ec,\rho_t}(\cdot )$ defined as in \eqref{rev} is the {\em time-reversal} of $\Ec$ for $\rho_t$, i.e.: 
\beq\label{revprop}\rho_t={\cal T}_{\rho_t}(\Ec^\dag)=\Rc^\dag_{\Ec,\rho_t}(\rho_{t+1}),\eeq and
\beq\label{consistency} {\cal T}_{\rho_{t+1}}(\Rc^\dag_{\Ec,\rho_t})(\sigma_t)=\Ec^\dag(\sigma_t),\eeq
for all $\sigma_t\in{\frak D}(\Hi)$ such that { $\supp(\sigma_t)\subseteq \supp(\rho_{t}).$ } 
Moreover, it can be augmented to be TPCP without affecting properties \eqref{revprop}-\eqref{consistency}. \footnote{By augmenting a Kraus map $\Ec$ with Kraus operators $\{ M_k \}_{k=1,\ldots, m}$ to a TPCP map, we mean adding a finite number $N$ of Kraus operators $\{ M_k \}_{k=m+1,\ldots, m+N}$ such that $\sum_k M^\dag_k M_k = I.$}
\end{theorem}

%

\begin{remark} Notice that if $\rho_t$ is full rank, \eqref{consistency} implies that ${\cal T}_{\rho_{t+1}}\circ{\cal T}_{\rho_{t}}$ is the the identity map on quantum operations, i.e. the time reversal of the time-reversal is the original forward map. One can also show that in general the time-reversal mechanism is not unique \cite{PT}, just as in the classical case (see the discussion following \eqref{relpq} in Section \ref{markovchains}). While studying quantum error correction problems \cite{knill-nearoptimal}, the same $\Rc^\dag_{\Ec,\rho}(\cdot)$ has been suggested by Barnum and Knill as a near-optimal correction operator. In their setting, $\Ec^\dag(\rho)$ is full-rank, and represents the output of a channel $\Ec$ with input state $\rho=\sum_j p_j\rho_j,$ a statistical mixture of some quantum codewords of interest to be recovered. It has also been proven there that $\Rc^\dag_{\Ec,\rho}(\cdot)$ is independent of the particular Kraus representation of $\Ec$.\end{remark}
 
\noindent Given a quantum Markov process, generated by $\rho_0$ and a sequence of TPCP maps $\{\Ec^\dag_t\}_{t\in[0, T-1]},$ a sequence of observables $\{Y_t\}_{t\in[0, T-1]}$ is said to be {\em space-time harmonic in reverse-time} with respect to the family $\{\Rc_{\Ec_t,\rho_t}\}_{t\in[0,T-1]}$ if:
\beq \label{qsthr} Y_{t+1}=\Rc_{\Ec_t,\rho_t}(Y_t), \eeq
extending Definition \ref{qharmonic} in analogy with the classical case.

\section{Maximum entropy problems for quantum Markov channels}\label{quantumMEP}

In the quantum case, the definition of a path-space for a Markov process is not obvious. Here, we build up quantum trajectories associating at each time an observable quantity and conditioning the state and the evolution to measurements of such observables. We next proceed by formulating and solving quantum versions of  problems MEP1-MEP2. The framework we develop leads to results that are in striking analogy with the classical case. The third case, where both marginals are prescribed, presents some intrinsic difficulties when one tries to generalize the classical approach, essentially due to the definition of quantum relative entropy.  A discussion of this problem is deferred to future publications. This issue is strictly related to the kinematical structure we highlight in \cite{PT}, where an alternative definition of quantum relative entropy emerges as a natural candidate.  Further work is also needed to address the possibility of extending the framework to include generalized or indirect measurement on the system, as it is typically the case in quantum optics \cite{carmicheal}. In our setting, anyway, commutativity of observables associated to different times is not required.

\subsection{Path spaces for quantum Markov processes}

Consider a quantum Markov process for a finite dimensional system $\cal Q$ with associated Hilbert space $\Hi_{\cal Q}$, generated by an initial density matrix $\sigma_0$ and a sequence of TPCP maps $\{\Ec^\dag_t\}_{t\in[0, T-1]},$ with each $\Ec^\dag_t$ admitting a Kraus representation with matrices $\{M_k(t)\}.$ 

We define a set of possible trajectories, or {\em quantum paths}, by considering a time-indexed family of observables $\{X_t\},\,X_t=\sum_{i=1}^{m_t} x_{i}\Pi_{i}(t),$ with $t\in[0,T].$ The paths are then all the possible time-ordered sequences of events  $\left(\Pi_{i_0}(0),\Pi_{i_1}(1),\ldots,\Pi_{i_T}(T)\right),$ with $i_t\in[1,\,m_t].$
By using \eqref{joinp}, we can compute the joint probability for a given path with the nested expression:
\beq \label{weight} w^\Ec_{(i_0,{i_1},\ldots,{i_T})}(\sigma_0) = \trace\left(\Pi_{i_T}(T)\Ec^\dag_{T-1}(\Pi_{i_{T-1}}(T-1)\ldots \Ec^\dag_{0}(\Pi_{i_0}(0)\sigma_0\Pi_{i_0}(0)) \ldots )\Pi_{i_T}(T)\right).\eeq


\begin{lemma} Define the {\em path-conditioned density matrices} for $t\in[0,T]$ via the relations
\beqa &&\hat\sigma_{\Ec,0}=\sum_{i_0}\Pi_{i_0}(0)\sigma_0\Pi_{i_0}(0),\nonumber\\
&&\hat\sigma_{\Ec,t+1}=\hat\Ec^\dag_t(\hat\sigma_{\Ec,t})=\sum_{i_{t+1}}\Pi_{i_{t+1}}(t+1)\Ec^\dag_{t}(\hat\sigma_{\Ec,t})\Pi_{i_{t+1}}(t+1),\label{conditionedstate}\eeqa 
where $\hat\Ec^\dag_t$ is TPCP and can be represented with double-indexed Kraus operators $\{\Pi_{i}(t+1)M_k(t)\}.$ The {\em marginal distribution} $w^\Ec_{i_t}(\sigma_0)$ at time $t\in[0,T],$ is then given by:
\beq w^\Ec_{i_t}(\sigma_0)=\trace\Big(\Pi_{i_t}(t)\hat\sigma_{\Ec,t}\Pi_{i_t}(t)\Big).\eeq

\end{lemma}
\proof From \eqref{weight}, by using the cyclic property of trace and the fact that $\Ec^\dag_T$ is trace-preserving, we get:
\beqan \sum_{i_T}w^\Ec_{(i_0,{i_1},\ldots,{i_T})}(\sigma_0) &=& \trace\left(\left(\sum_{i_T}\Pi_{i_T}(T)\right)\Ec^\dag_{T-1}(\Pi_{i_{T-1}}(T-1)\ldots \Ec^\dag_{0}(\Pi_{i_0}(0)\sigma_0\Pi_{i_0}(0)) \ldots )\right)\\
&=& w^\Ec_{(i_0,{i_1},\ldots,{i_{T-1}})}(\sigma_0).\eeqan
Hence, by iterating and substituting the definition \eqref{conditionedstate}:
\beqan w^\Ec_{i_t}(\sigma_0)&=&\sum_{i_0,\ldots,i_{t-1},i_{t+1},\ldots,{i_T}} w^\Ec_{(i_0,{i_1},\ldots,{i_T})}(\sigma_0)= \sum_{i_0,\ldots,i_{t-1}} w^\Ec_{(i_0,{i_1},\ldots,{i_t})}(\sigma_0)\\
&=&\trace\Big(\Pi_{i_t}(t)\Ec^\dag_{t-1}\Big(\sum_{i_{t-1}}\Pi_{i_{t-1}}(t-1)\ldots \Ec^\dag_{0}\Big(\sum_{i_0}\Pi_{i_0}(0)\sigma_0\Pi_{i_0}(0)\Big) \ldots \Pi_{i_{t-1}}(t-1)\Big)\Pi_{i_t}(t)\Big)\\
&=&\trace\Big(\Pi_{i_t}(t)\hat\sigma_{\Ec,t}\Pi_{i_t}(t)\Big).\eeqan.
\qed
\begin{remark} If for each $t$ $[X_t,\sigma_t]:=X_t\sigma_t-\sigma_tX_t=0$, then, computing the reduced density operators at each time, we obtain the original quantum Markov process, that is $\hat\sigma_t=\sigma_t$ for all $t$. In general this is not the case: Imposing a (finite) set of possible trajectories by choosing the $\{X_t\}$, we have to condition the density matrix at time $t$ on the past measurements. Unlike classical probability, even ``non-selective'' conditioning influences the state.  
\end{remark}
Observe moreover the following fact:
\begin{proposition} 
The joint probabilities can be re-written in terms of the time reversal transitions for the {\em path-conditioned states} as:
\beq \label{weightrev} w^{\Ec}_{(i_0,{i_1},\ldots,{i_T})}(\sigma_0) = \trace\left(\Pi_{i_0}(0)\Rc^\dag_{{\hat\Ec}_0,\hat\sigma_{\Ec,0}}(\Pi_{i_{1}}(1)\ldots \Rc^\dag_{{\hat\Ec}_{T-1},\hat\sigma_{{\Ec},T-1}}(\Pi_{i_T}(T)\hat\sigma_{{\Ec},T}\Pi_{i_T}(T)) \ldots )\Pi_{i_0}(0)\right).\eeq
\end{proposition}
\proof
Since  $[\hat\sigma_t,X_t]=0,\, \forall t\in[0,T],$ by using \eqref{backwardp} we have that:  
$$ w^{\Ec}_{(i_0,{i_1},\ldots,{i_T})}(\hat\sigma_{\Ec,0}) = \trace\left(\Pi_{i_0}(0)\Rc^\dag_{{\hat\Ec}_0,\hat\sigma_{\Ec,0}}(\Pi_{i_{1}}(1)\ldots \Rc^\dag_{{\hat\Ec}_{T-1},\hat\sigma_{{\Ec},{T-1}}}(\Pi_{i_T}(T)\hat\sigma_{{\Ec},T}\Pi_{i_T}(T)) \ldots )\Pi_{i_0}(0)\right).$$
By noting that $ w^{\Ec}_{(i_0,{i_1},\ldots,{i_T})}(\sigma_0)= w^{\Ec}_{(i_0,{i_1},\ldots,{i_T})}(\hat\sigma_{\Ec,0}),$ we get the conclusion.
\qed
\noindent This ``backward'' representation will play a key role in the solution of the maximum entropy problems we discuss in the next Section.

\subsection{Maximum entropy problems on quantum path spaces}
We consider the simpler maximum entropy problems where only the initial or final density matrices are prescribed. The solution to these problems exhibit the same structure of their classical analogues, involving  a ``symmetrized'' multiplicative functional transformation.

Let $\{\Ec^\dag_t\}$ be a family of TPCP maps generating a quantum Markov process over $[0,T]$ with initial density matrix $\sigma_0$. Assume that at time $T$ the density matrix of the system has been found to be $\bar\rho_T,$ being different from the expected $\sigma_T=\Ec^\dag_{T-1}\circ\ldots\circ\Ec^\dag_{0}(\sigma_0).$ Let $\{X_t\}$ be a time-indexed family of observables defining a path space as above. We constraint {\em only} $X_T$ to be such that $[X_T,\bar\rho_T]=0,$ and it admits a spectral decomposition with rank one $\Pi_j(T)$'s (this is quite natural, since $\bar\rho_T$ is given). Let as above $w^\Ec(\sigma_0)$ denote the path-space distribution induced by the initial condition $\sigma_0$ and the TPCP transitions $\{\Ec^\dag_t\}$. For simplicity, in the reminder of the section, the reverse-time quantum operations are assumed to be trace preserving. The general case is simply obtained by augmenting the Kraus operators in order to have a trace preserving transformation, as detailed in Section \ref{timerev}.
Consider now the \\{\em Quantum Maximum Entropy Problem} (QMEP1): 
\begin{equation}\label{QMEs1}{\rm minimize}\quad \left\{\D(w^\Fc(\rho_0)\|w^\Ec(\sigma_0)); w^\Fc(\rho_0)\in \Omega(\bar\rho_T) \right\}\end{equation}
with $\Omega(\bar\rho_T)$ the set of path space distribution induced by a quantum Markov process generated by a family of TPCP maps $\{\Fc^\dag_t\}$ and some initial $\rho_0$ such that their path-conditioned, final density matrix satisfies $\hat\rho_{\Fc,T}=\bar\rho_T$.

\noindent In order to solve Problem QMEP1, notice the following facts.
\begin{enumerate}

\item Since we required the $\Pi_i(T)$'s to be rank-one, it follows that for all $i\in[1,m_T]$:
$$\Pi_{i}(T)\rho_T\Pi_{i}(T)=\trace\big(\rho_T\Pi_{i}(T)\big)\Pi_{i}(T)=w^\Fc_{i_T}(\rho_0)\Pi_{i}(T).$$

\noindent Hence, we can write
\beq\label{qback}w^\Fc_{(i_0,{i_1},\ldots,{i_T})}(\rho_0)=w^\Fc_{(i_0,{i_1},\ldots,{i_{T-1}}|i_T)}\cdot w^\Fc_{i_T}(\rho_0),\eeq
\noindent where we used the expression \eqref{weightrev} and defined the {\em conditional probabilities}:
\beq\label{qconditioned}w^\Fc_{(i_0,{i_1},\ldots,{i_{T-1}}|i_T)}=\trace\left(\Pi_{i_0}(0)\Rc^\dag_{{\hat\Fc}_0,\hat\rho_0}(\Pi_{i_{1}}(1)\ldots \Rc^\dag_{{\hat\Fc}_{T-1},\hat\rho_{T-1}}(\Pi_{i_T}(T)) \ldots )\Pi_{i_0}(0)\right).\eeq\end{enumerate}
Decomposition \eqref{qback} clearly plays the role of expressions \eqref{backrec1}-\eqref{backrec2} in the classical case.
By employing \eqref{qback} and its equivalent for $w^\Ec_{(i_0,{i_1},\ldots,{i_T})}(\sigma_0),$
we can now write:

\beqa \nonumber\D(w^\Fc_{(i_0,i_1,\ldots,i_T)}(\rho_0)\|w^\Ec_{(i_0,{i_1},\ldots,{i_T})})(\sigma_0))&=& \sum_{i_0,{i_1},\ldots,{i_T}}  w^\Fc_{(i_0,{i_1},\ldots,{i_T})}(\rho_0)\log\frac{ w^\Fc_{(i_0,{i_1},\ldots,{i_T})}(\rho_0)}{w^\Ec_{(i_0,{i_1},\ldots,{i_T})}(\sigma_0)}\\
&=&\sum_{i_T}\left(\sum_{i_0,\ldots,{i_{T-1}}}  w^\Fc_{(i_0,{i_1},\ldots,{i_{T-1}}|i_T)}\log\frac{ w^\Fc_{(i_0,{i_1},\ldots,{i_{T-1}}|i_T)}}{w^\Ec_{(i_0,i_1,\ldots,{i_{T-1}}|i_{T})}}\right) w_{i_T}^\Fc(\rho_0)\nonumber\\&& +\;\sum_{i_T}w^\Fc_{{i_T}}(\rho_0)\log\frac{w^\Fc_{{i_T}}(\rho_0)}{w^\Ec_{{i_T}}(\sigma_0)}.\nonumber\\ \label{decomp}
\eeqa

\noindent Since $\hat\rho_{\Fc,T}=\sum_{i_T}w^\Fc_{{i_T}}(\rho_0)\Pi_{i_T}=\bar\rho_T,$ notice that the first part of \eqref{decomp}  is non-negative and independent of the initial state, while the second is independent of the choice of ${\Omega^\Fc(\rho_0)\in\mathfrak D}(\bar\rho_T)$. If we can obtain a process with conditioned probabilities such that: 
$$ { w^\Fc_{(i_0,{i_1},\ldots,{i_{T-1}}|i_T)}}={w^\Ec_{(i_0,i_1,\ldots,{i_{T-1}}|i_T)}}, $$
the first part of \eqref{decomp} becomes zero, providing us with an optimal solution. By \eqref{qconditioned}, a sufficient condition for optimality is to have the {\em same path-conditioned backward mechanism} as the reference evolution. We have thus proven the following quantum counterpart of Theorem \ref{thmminentr}:
\begin{theorem}\label{qtheo1}A solution to (QMEP1) \eqref{QMEs1} is given by the quantum Markov process with path-conditioned final density $\bar\rho_T$ at time $T$ and reverse-time transition mechanism equal to that of $\{\Ec_t\}$, namely
\begin{equation} \Rc^\dag_{\hat\Fc_t,\hat\rho_{\Fc,t}}(\cdot)=\Rc^\dag_{\hat\Ec,\hat\sigma_{\Ec,t}}(\cdot ),\quad\forall t\in[0,T-1].
\end{equation}
\end{theorem}

Notice that with this optimal choice, the total cost is bounded by the relative entropy of the conditioned final density matrices: 
\beq\label{QMEP1cost} \sum_{i_T}w^\Fc_{{i_T}}(\rho_0)\log\frac{w^\Fc_{{i_T}}(\rho_0)}{w^\Ec_{{i_T}}(\sigma_0)}=\D(\bar\rho_T\|\hat\sigma_{\Ec,T}).\eeq
Let us compute the ``forward'' quantum operations, which, as in the classical case, will turn out to be time dependent even when the reference process is time-homogeneous. By Theorem \ref{qtimerev}, recalling that the conditioned transition mechanism $\hat\Ec^\dag_t$ admits a Kraus representation with operators $\Pi_j(t+1)M_k(t)$, see \eqref{conditionedstate}, one finds that the Kraus operators of $\Rc^\dag_{\Ec_t,\hat\sigma_{\Ec,t}}$ are given by the double-indexed $R_{j,k}(\hat\Ec_t,\hat\sigma_{\Ec,t})=\hat\sigma_{\Ec,t}^\um M^\dag_k(t)\Pi_j(t+1) \hat\sigma_{\Ec,t+1}^{-\um}.$ Reversing this TPCP map, now with respect to the state $\hat\rho_{\Fc,t+1},$ we get:
$$F_{j,k}(t)=\hat\rho_{\Fc,t+1}^{\um}\hat\sigma_{\Ec,t+1}^{-\um} \big(\Pi_j(t+1)M_k(t)\big) \hat\sigma_{\Ec,t}^{\um}\hat\rho_{\Fc,t}^{-\um}.$$
which can be consider as a {\em non-commutative, ``symmetrized''} version of a multiplicative functional transformation in the classical case. In fact, define $N_t=\hat\rho^{\um}_{\Fc,t}\hat\sigma^{-\um}_{\Ec,t}.$ Then we have that the TPCP maps of the solution, with initial condition $\hat\rho_{\Fc,0},$ are given by: $$\Fc^\dag_t(\rho)= \sum_{j,k}  N_{t+1}\big(\Pi_j(t+1)M_k(t)\big) N^{-1}_t\,\rho \,  N^{-\dag}_t \big(M_k^\dag(t)\Pi_j(t+1) \big) \,N_{t+1}^\dag,$$
which has the role of the transitions defined in \eqref{multiplfunct}.
\noindent Moreover, $Y_t=N_t^\dag N_t=\hat\sigma_{\Ec,t}^{-\um}\hat\rho_{\Fc,t}\hat\sigma_{\Ec,t}^{-\um}$ satisfies:
\beqan\hat\Ec_t(Y_{t+1})&=&\sum_{j,k} \Pi_j(t+1)M_k^\dag(t) \hat\sigma_{\Ec,t+1}^{-\um}\hat\rho_{\Fc,t+1}\hat\sigma_{\Ec,t+1}^{-\um} M_k(t)\Pi_j(t+1)\\&=&\sum_{j,k} \hat\sigma_{\Ec,t}^{-\um} R_{j,k}(\Ec_t,\hat\sigma_{\Ec,t})\hat\rho_{\Fc,t+1}R_{j,k}^\dag(\Ec_t,\hat\sigma_{\Ec,t})\hat\sigma_{\Ec,t}^{-\um} =Y_t,\eeqan
that is, $\{Y_t\}$ is {\em space-time harmonic} with respect to the transition $\hat\Ec_t,$ completing the  analogy to \eqref{multiplfunct}. We remark that, since every time-reversal can be augmented to be TPCP by Theorem \ref{qtimerev}, one can always complete $\Rc^\dag_{\hat\Ec_t,\hat\sigma_{\Ec,t}}(\cdot ),$ and then $\Fc^\dag_t(\cdot ),$ to be TPCP.

Consider now the case where the {\em initial state} is constrained to be equal to $\bar\rho_0,$ different from the a-priori initial condition $\sigma_0$. Consider a path-space induced by observables $\{X_t\}$ such that $X_0$ has non-degenerate spectrum. By using \eqref{weight}, we can write:
\beqan w^\Ec_{(i_0,{i_1},\ldots,{i_T})}(\sigma_0) &=& \trace\left(\Pi_{i_T}(T)\Ec^\dag_{T-1}(\Pi_{i_{T-1}}(T-1)\ldots \Ec^\dag_{0}(\Pi_{i_0}(0)) \ldots )\Pi_{i_T}(T)\right)\cdot w^\Ec_{(i_0)}(\sigma_0)\\
&=& w^\Ec_{({i_1},\ldots,{i_T}|i_0)}\cdot w^\Ec_{(i_0)}(\sigma_0).\eeqan
Arguing as above, we get:
\beqa \nonumber\D(w^\Fc_{(i_0,i_1,\ldots,i_T)}(\bar\rho_0)\|w^\Ec_{(i_0,{i_1},\ldots,{i_T})})(\sigma_0))
&=&\sum_{i_0}\left(\sum_{i_1,\ldots,i_{T}}  w^\Fc_{({i_1},\ldots,{i_T)}|i_0)}\log\frac{ w^\Fc_{({i_1},\ldots,{i_T)}|i_0)}}{w^\Ec_{({i_1},\ldots,{i_T)}|i_0)}}\right) w_{i_0}^\Fc(\bar\rho_0)\nonumber\\&&+\;\sum_{i_T}w^\Fc_{{i_0}}(\bar\rho_0)\log\frac{w^\Fc_{{i_0}}(\bar\rho_0)}{w^\Ec_{{i_0}}(\sigma_0)}\nonumber\\ \label{decomp2}
\eeqa

\begin{theorem}\label{qtheo2}A solution to (QMEP2)\\
\begin{equation}\label{QMEs2}{\rm minimize}\quad \left\{\D(w^\Fc(\bar\rho_0)\|w^\Ec(\sigma_0)); w^\Fc(\bar\rho_0)\in {\Omega}(\bar\rho_0) \right\}\end{equation}
with ${\Omega}(\bar\rho_0)$ the set of path space probability distributions induced by a family of TPCP maps $\{\Fc^\dag_t\}$ and initial state $\bar\rho_0,$
is given by the quantum Markov process with initial density $\bar\rho_0$ and forward transitions: 
\beq\Fc_t(\cdot)=\Ec_t(\cdot),\quad\forall t\in[0,T-1].
\eeq
\end{theorem}
\begin{remark} 
Altough the QMEP2 problem apparently depends on the choice of the quantum path-space, that is the observables $\{X_t\}_{t\in[0,T]},$ we remark that {\em its solution does not.} The difference between problems QMEP1 and QMEP2 is given by the fact that in QMEP2 we are concerned with the forward transitions, and we do not need to use the path-conditioned density matrices \eqref{conditionedstate}. The classical case does not present this asymmetry since classical non-selective measurements do not alter the state. \end{remark}

The final cost admits a bound similar to that in Problem QMEP1, that can be easily related to the unconditioned states. In fact, using monotonicity of relative entropy with respect to conditioning \cite{lindblad-entropy}, we get:
\beq\sum_{i_T}w^\Fc_{{i_0}}(\bar\rho_0)\log\frac{w^\Fc_{{i_0}}(\bar\rho_0)}{w^\Ec_{{i_0}}(\sigma_0)}=\D(\hat\rho_0\|\hat\sigma_{\Ec,0})=\D(\bar\Ec^\dag(\bar\rho_0)\|\bar\Ec^\dag(\sigma_0))\leq\D(\bar\rho_0\|\sigma_0),\eeq
with $\bar\Ec^\dag(\rho)=\sum_i\Pi_i(0)\rho\Pi_i(0).$

Notice that the operator-sum of the two reverse-time evolutions $\Rc_{\Fc,\rho_t},\Rc_{\Ec,\sigma_t}$ satisfy, under appropriate restriction on the support of $\rho_t,\sigma_t$: $$R_k(\Fc_t,\rho_t)=\rho_t^{\um}M_k^\dag(t) \rho_{t+1}^{-\um}= \rho_t^{\um}\sigma_t^{-\um}R_k(\Ec_t,\sigma_t)\sigma_{t+1}^{\um}\rho_{t+1}^{-\um},$$
which is again as a quantum symmetrized ``multiplicative'' functional transformation.
To carry on the analogy with the classical case, consider two quantum Markov processes corresponding to different initial conditions $\rho_0\neq\sigma_0$ but with same family of trace-preserving quantum operations $\{\Ec^\dag_t\}$. Assume for simplicity $\supp(\rho_t)\subset\supp (\sigma_t),\,\forall t.$ Define the observable $Y_t=\sigma_t^{-\um}\rho_t\sigma_t^{-\um}.$
We thus have that:
\begin{equation}\label{rtspacetimeq}\Rc_{\Ec,\sigma_t}(Y_t)=\sum_k \sigma_{t+1}^{-\um} M_k(t) \sigma_t^{\um} \sigma_t^{-\um}\rho_t\sigma_t^{-\um}\sigma_t^{\um}M_k^\dag (t) \sigma_{t+1}^{-\um}=Y_{t+1}.
\end{equation}
This shows that $Y_t$ evolves in the forward direction of time with the backward mechanism of $\sigma_t,$. It is namely quantum {\em space-time harmonic in reverse time} with respect to the transition of $\sigma_t$, once more in striking analogy with \eqref{rtspacetime}. Moreover, the special case  of property (\ref{rtspacetimeq}) where the quantum operation $\Ec^\dag$ does not depend on time and $\sigma_t\equiv\bar{\sigma}$ is an invariant density matrix for  $\Ec^\dag$ is, as in the classical case, connected to a strong form of the second law \cite{PT}.

\section{Conclusion and outlook}

In this paper, we have studied a class of maximum entropy problems on path space first for Markov chains and then for quantum channels. In both cases, solutions are obtained from the ``prior" evolution via a multiplicative functional transformation induced by a space-time harmonic function in analogy to what was known for the diffusion case. The classical theory of Schr\"{o}dinger bridges is connected to a variety of other fascinating topics besides large deviations. First of all, there is Schr\"{o}dinger's original motivation: He had observed the strong analogy between the time reversibility of the solution bridge and that of quantum mechanics\footnote{``Merkw\"{u}rdige Analogien zur
Quantenmechanik, die mir sehr  des Hindenkens wert erscheinen".}. This motivation is from time to time rediscovered in various versions of stochastic mechanics see e.g. \cite{Z,nag,LK2,P}. Other related topics are reciprocal processes ($1D$ Markov fields), see e.g. \cite{Jam,LK1,LK2}, stochastic control and variational principles \cite{W,pavwak}, logarithmic transformations of parabolic equations \cite{FS}, Feynman-Kac formula \cite{KS,PDM}, etc.. In order to avoid overburdening this first manuscript, we have chosen to defer discussing/developing these connections for Markov chains to future publications.
There is, however, another link: The reverse time space-harmonic functions occurring in the solutions of problems (MEP2) and (QMEP2) lead to a strong form of the second law. This is presented in \cite{PT}.

The  Markov chain Schr\"{o}dinger bridges appear as a flexible tool to be tested on a variety of applications, given the recent surfacing of the full modeling and computational power of Markov chains, cf. e.g. \cite{bremaud, hagg,PDM,mitupf}.
For quantum systems, this framework may be useful to attack steering problems \cite{BFP} and to complement or improve quantum process tomography techinques (see e.g. \cite{lidar-tomography} for a recent review of different methods). Exploring the relations of our framework with the theory of quantum error correction \cite{knill-qec, nielsen-chuang, knill-nearoptimal, viola-generalnoise} appears to be a particularly promising research direction. The problem of finding the time-reversal of quantum operations or quantum Markov semigroups  \cite{alicki-lendi,breuer-petruccione} representing the effect of noisy channels on some quantum code is strictly related to many central problems in quantum information  and its realizations. Moreover, our path-space problems appear to be compatible with the general setting proposed in \cite{BDKSSS} to develop a quantum version of Sanov's theorem for product states. This suggests that our results may play a role in 
hypothesis testing and large deviation theory  for quantum Markov evolution, once more in remarkable analogy with the classical setting. 


\section*{Acknowledgment}
The authors wish to thank Lorenza Viola for valuable discussion on quantum measurements, trajectories and error correction, and for her interest in this work.

\appendix
\section{Discrete-time martingales}\label{dismart}
A general reference is \cite{neveu}. Following \cite{bremaud}, we give a somewhat restricted definition of martingale.
\begin{definition}\label{defmart}Consider a discrete time stochastic process $X=\{X(t), t\ge 0\}$ with finite or denumerable state space ${\cal X}$. The process $Y=\{Y(t), t\ge 0\}$ is called a {\em martingale with respect to} $X$ if
\begin{enumerate}
\item $\E|Y(t)|<\infty,\forall t\ge 0$;
\item $Y(t)$ is a function of $\{X(s),0\le s\le t\},\forall t\ge 0$;
\item $\E(Y(t+1)|X(0),X(1),\ldots, X(t))=Y(t)$.
\end{enumerate}
\end{definition}
We can say that a martingale is {\em conditionally constant}. Notice that the case $Y(t)=X(t)$ is also included.
An elementary example of a martingale is provided by the capital of a player at time $t$ in a fair coin tossing game.
\begin{proposition}\label{spacemart} Let $h$ be space-time harmonic for the  Markov chain $X=\{X(t); t\in\N\}$ with state space ${\cal X}$ and transition matrix $P(t)=\left(p_{ij}(t)\right)$. Define the stochastic process $Y=\{Y(t)=h(t,X(t)), t\ge 0\}$. Then, if $E|Y(t)|<\infty,\forall t$, $Y$ is a martingale with respect to $X$. 
\end{proposition}
\proof
We have
\begin{eqnarray}\nonumber \E(Y(t+1)|X(0),X(1),\ldots, X(t))&=&\E(h(t+1,X(t+1))|X(0),X(1),\ldots, X(t))\\&=&\E(h(t+1,X(t+1))| X(t)).\nonumber
\end{eqnarray}
Now observe that for all $i\in{\cal X}$
$$\E(h(t+1,X(t+1))| X(t)=i)=\sum_{j}p_{ij}(t)h(t+1,j)=h(t,i).
$$
Thus $\E(h(t+1,X(t+1))| X(t))=h(t,X(t))=Y(t)$. Since properties $1)$ and $2)$ are also satisfied, we conclude that $Y$ is a martingale with respect to $X$.
\qed
\section{Proof of Lemma \ref{lemmarelentr}.}\label{prooflemma}
Using (\ref{backrec1})-(\ref{backrec2}), we get
\begin{eqnarray}\nonumber\D({\bf P}\|{\bf \Pi})=\sum_{i_0}\sum_{i_1}\cdots\sum_{i_{T}}{\bf P}(i_0,i_1,\ldots,i_{T-1},i_T)\log\frac{{\bf P}(i_0,i_1,\ldots,i_{T-1},i_T)}{{\bf \Pi}(i_0,i_1,\ldots,i_{T-1},i_T)}\\\nonumber=\sum_{i_o,i_1,\cdots,i_T}{\bf P}(i_0,i_1,\ldots,i_{T-1},i_T)\left[\sum_{k=1}^T\log\left(\frac{q_{i_ki_{k-1}}(k-1)}{q^\pi_{i_ki_{k-1}}(k-1)}\right)\,+\log\left(\frac{p^1_{i_T}}{\pi_{i_T}(T)}\right)\right].
\end{eqnarray}
Observe now that
\begin{eqnarray}\nonumber \sum_{i_o,i_1,\cdots,i_T}{\bf P}(i_0,i_1,\ldots,i_{T-1},i_T)\log\left(\frac{p^1_{i_T}}{\pi_{i_T}(T)}\right)\\\nonumber=\sum_{i_T}\sum_{i_o,i_1,\cdots,i_{T-1}}{\bf P}(i_0,i_1,\ldots,i_{T-1},i_T)\log\left(\frac{p^1_{i_T}}{\pi_{i_T}(T)}\right)\\\nonumber=\sum_{i_T}p^1_{i_T}\log\left(\frac{p^1_{i_T}}{\pi_{i_T}(T)}\right)=\D\left(p^1\|\pi(T)\right).
\end{eqnarray}
Moreover, let $p(i_{k-1},i_k)=\Prob(X(k-1)=i_{k-1},X(k)=i_k)$ be the two times marginal of ${\bf P}$. Observe that $p(i_{k-1},i_k)=q_{i_ki_{k-1}}(k-1)p_{i_k}(k)$. We then get
\begin{eqnarray}\nonumber \sum_{i_o,i_1,\cdots,i_T}{\bf P}(i_0,i_1,\ldots,i_{T-1},i_T)\sum_{k=1}^T\log\left(\frac{q_{i_ki_{k-1}}(k-1)}{q^\pi_{i_ki_{k-1}}(k-1)}\right)\\\nonumber=\sum_{k=1}^T\sum_{i_k,i_{k-1}}\sum_{i_j\neq i_k, i_{k-1}}{\bf P}(i_0,i_1,\ldots,i_{T-1},i_T)\log\left(\frac{q_{i_ki_{k-1}}(k-1)}{q^\pi_{i_ki_{k-1}}(k-1)}\right)\\\nonumber=\sum_{k=1}^T\sum_{i_k,i_{k-1}}p(i_{k-1},i_k)\log\left(\frac{q_{i_ki_{k-1}}(k-1)}{q^\pi_{i_ki_{k-1}}(k-1)}\right)\\\nonumber =\sum_{k=1}^T\sum_{i_k,i_{k-1}}q_{i_ki_{k-1}}(k-1)p_{i_k}(k)\log\left(\frac{q_{i_ki_{k-1}}(k-1)}{q^\pi_{i_ki_{k-1}}(k-1)}\right)\\\nonumber=\sum_{k=1}^T\sum_{i_k}\D\left(q_{i_ki_{k-1}}(k-1)\|q^\pi_{i_ki_{k-1}}(k-1)\right)p_{i_k}(k)
\end{eqnarray}
and (\ref{relentrdec}) follows.
\qed
\section{Application to large deviations}\label{LD}
The area of large deviations is concerned with the probabilities of very rare events. Let $X_1,X_2,\ldots$ be i.i.d. random variables with finite expectation. The strong law of large numbers asserts that
$$\frac{1}{N}\sum_{i=1}^NX_i\rightarrow \E(X_1), \quad {\rm a.s.}.
$$
The study of deviations from the typical behavior is central in probability. We consider first {\em normal deviations} (called fluctuations by physicists). The central limit theorem implies that the quantity
$$\Prob(|\frac{1}{N}\sum_{i=1}^NX_i-\E(X_1)|\ge\frac{\epsilon}{\sqrt{N}})
$$
converges as $N$ tends to infinity to a positive value. In (level-1) {\em large deviations} one considers instead the asymptotic behavior of
$$\Prob(|\frac{1}{N}\sum_{i=1}^NX_i-\E(X_1)|\ge\epsilon.
$$
By the law of large numbers, this quantity tends to zero. But one is interested in learning if it decays exponentially (it does) and if so with what exponent. We refer to the treatises \cite{ellis,DS,DZ} for a thorough introduction and to \cite{DGW,Wak} for large deviations in connections to classical Schr\"{o}dinger bridges. We mention, in passing, that large deviations theory has various  applications in hypothesis testing, rate distortion theory, etc, see e.g. \cite[Chapter 11]{cover}, \cite[Chapters 2,3,7]{DZ}. We now investigate the application of the solution to (MEP3) to  large deviations of the empirical distribution (level-2 large deviations). 

Let ${\cal X}$ be a finite or countably infinite set. Let $X^1,X^2,\ldots$ be independent, identically distributed Markov evolutions on the discrete time interval $[0,T]$ with state space ${\cal X}$  and defined on the probability space $(\Omega, {\cal F}, \Prob)$. Let ${\bf \Pi}$ be their common distribution on ${\cal X}^{T+1}$. We can alternatively think of the Markov chains $X^i$ as random variables  taking values in the $\sigma$-compact metric space $({\cal X}^{T+1},d)$ where $d$ denotes the discrete metric. The {\em empirical distribution} $\mu_n$ associated to $X^1,X^2,\ldots X^n$ is defined by
\begin{equation}\label{emp}\mu_n(\omega):=\frac{1}{n}\sum_{i=1}^n\delta_{X^i}(\omega),\quad \omega\in\Omega.
\end{equation}
Notice that (\ref{emp}) defines a map from $\Omega$ to the space ${\cal D}({\cal X}^{T+1})$ of probability distributions on ${\cal X}^{T+1}$. Hence, if $E$ is a subset of ${\cal D}({\cal X}^{T+1})$, it makes sense to consider $\Prob(\omega: \mu_n(\omega)\in E)$. By the ergodic theorem, see e.g. \cite[Theorem A.9.3.]{ellis}, the distributions $\mu_n$ converge weakly 
\footnote{Let ${\cal V}$ be a metric space and ${\cal D}({\cal V})$ be the set of probability measures defined on ${\cal B}({\cal V})$, the Borel $\sigma$-field of ${\cal V}$. We say that a sequence $\{P_n\}$ of elements of ${\cal D}({\cal V})$ converges weakly to $P\in {\cal D}({\cal V})$, and write  $P_n\Rightarrow P$, if $\int_{\cal V}fdP_n\rightarrow\int_{\cal V}dP$ for every bounded, continuous function $f$ on ${\cal V}$.} to ${\bf \Pi}$ as $n$ tends to infinity. Hence, if ${\bf \Pi}\not\in E$, we must have $\Prob(\omega: \mu_n(\omega)\in E)\searrow 0$. Large deviation theory provides us with a much finer result: Such a decay is {exponential} and the exponent may be characterized solving a maximum entropy problem! Indeed, in our setting, let $E={\cal D}(0,T;p^0,p^1)$. Then, Sanov's theorem \cite[Theorem 8.2]{azen}, roughly asserts that if the ``prior" ${\bf \Pi}$ does not have the required marginals, the probability of observing an empirical distribution $\mu_n$ in ${\cal D}(0,T;p^0,p^1)$ decays according to 

$$\Prob\left(\frac{1}{n}\sum_{i=1}^n\delta_{X^i}\in{\cal D}(0,T;p^0,p^1)\right)\sim\exp\left[-n \inf\left\{\D({\bf P}\|{\bf \Pi}); {\bf P}\in{\cal D}(0,T;p^0,p^1)\right\} \right]. 
$$
In the words of \cite{DZ}: ``Sanov's theorem provides a quite unexpected link between Large Deviations, Statistical Mechanics, and Information Theory". Let us go back to the solution of problem (MEP3). Since the optimal distribution $\hat{\bf P}$ of Theorem \ref{maincl} makes the cost function $J({\bf P})$ of Problem \ref{ME5} equal to  zero, we get that the exponent is given by
\begin{eqnarray}\nonumber\D(\hat{{\bf P}}\|{\bf \Pi})=\D(p^0\|\pi(0))-\sum_{i_0}\log\varphi (0,i_0)p^0_{i_0}+\sum_{i_T}\log\varphi(T,i_T)p^1_{i_T}\\=\sum_{i_0}\left[\log\frac{p^0_{i_0}}{\pi_{i_0}(0)\varphi (0,i_0)}\right]p^0_{i_0}+\sum_{i_T}\log\varphi(T,i_T)p^1_{i_T},\nonumber
\end{eqnarray}
where $\varphi$ is computed through the Schr\"{o}dinger system (\ref{scsi1})-(\ref{scsi2})-(\ref{scsi3}).
There are, of course, similar large deviations results involving Problems (MEP1) and (MEP2). In the first case, where the final marginal is fixed, the exponent is given by $\D(p^1\|\pi(T))$. Symmetrically, in the second case, the exponent is $\D(p^0\|\pi(0))$.

\end{document}